\documentclass[12pt]{article}
\usepackage{epsfig}

\title{\large \bf  Spontaneous violation of chiral symmetry \\
in QCD vacuum is the origin of baryon masses\\ and determines baryon magnetic moments
and\\ their other static properties}
\author{B.L.Ioffe\\ \\
A.I.Alikhanov Institute of Theoretical and \\ Experimental Physics, Moscow, Russia}
\begin{document}
\date{}
\maketitle

\newcommand{\be}{\begin{equation}}
\newcommand{\ee}{\end{equation}}

\def\la{\mathrel{\mathpalette\fun <}}
\def\ga{\mathrel{\mathpalette\fun >}}
\def\fun#1#2{\lower3.6pt\vbox{\baselineskip0pt\lineskip.9pt
\ialign{$\mathsurround=0pt#1\hfil##\hfil$\crcr#2\crcr\sim\crcr}}}

\begin{abstract}

A short review is presented of the spontaneous violation of chiral symmetry  in QCD
vacuum. It is demonstrated, that this phenomenon is the origin of baryon masses in QCD.
The value of nucleon mass is calculated as well as the masses of hyperons and some
baryonic  resonances and expressed mainly through the values of quark condensates --
$\langle 0\mid \bar{q} q \mid 0\rangle, ~q=u,d,s$ -- the vacuum expectation values
(v.e.v.) of quark field. The concept of vacuum expectation values induced by external
fields is introduced. It is demonstrated that such v.e.v. induced by static
electromagnetic field results in quark condensate magnetic susceptibility, which  plays
the main role in determination of baryon magnetic moments. The magnetic moments of
proton, neutron and hyperons are calculated. The results of calculation of baryon octet
$\beta$-decay constants are also presented.

\end{abstract}


\vspace{1cm}

This paper is  dedicated to the 85-th anniversary of Spartak Timofeevich Belyaev. Among
all of his striking qualities I would like to especially stress the following  -- his
principiality, the  principiality both in science, and in social life. Spartak (I tell
him so, because of our old friendship) is not going to any compromise, leaves not a
position, which he considers  as principal, although he understands  well, that this will
result in serious troubles to him. For this quality (but not only for it !), I have a
great respect to him and wish him to be such for long years.

\newpage

{\bf \large 1. Chiral symmetry of QCD and its spontaneous violation}

\bigskip

 As is well known [1,2] the masses of light $u,d,s$ quarks, which enter the QCD
 Lagrangian, especially the masses of $u$ and $d$ quarks from which the usual
 (nonstrange) hadrons are built, are very small as compared to the characteristic  QCD
 mass scale  $M_{char}\sim 1$ GeV:
 \be
 \frac{m_u}{M_{char}} < 0.01,~~~\frac{m_d}{M_{char}} < 0.01,
 ~~~\frac{m_s}{M_{char}}\approx 0.15\label{1}\ee In QCD the quark interaction is due to
 the exchange of vector gluonic field. Thus, if light quark masses are neglected, the QCD
 Lagrangian (its light quark part) becomes chirally  invariant -- the quark fields can be
 transformed as $q\to \gamma_5 q$ or $q \to e^{i\alpha\gamma_5}q, ~q=u,d,s$. That means
 that not only vector, but also axial currents are  conserved. (With exception of singlet
 (in flavour) axial current, which is not conserved because of anomaly -- see e.g.
 [3].) However, the chiral symmetry is not realized in the spectrum of hadrons and
 their low energy interactions. Particularly, it can be shown, that v.e.v. $\langle 0\mid
 \bar{q} q \mid 0\rangle$, which should be equal to zero, if chiral symmetry is fulfilled, is
 not zero in fact:
 \be
 \langle 0 \mid \bar{q} q   \mid 0 \rangle_{1 GeV} =
 -\frac{1}{2}\frac{m^2_{\pi}f^2_{\pi}}{m_u +m_d} \approx -(254~\mbox{MeV})^3,
 ~~~q=u,d\label{2}\ee -- the Gell-Mann, Oakes, Renner theorem [4] (for the proof see
 [5]). Here $m_{\pi}$ is the pion mass, $f_{\pi}$ is the pion decay constant,
 $f_{\pi} = 130.7$ MeV [6]) $m_u, m_d$ -- are $u$- and $d$- quark masses ($m_u
 \approx 3.0$ MeV, $m_d \approx 7.0$ MeV  at the normalization point $\mu =1$ GeV
 [7].) In a chirally symmetric theory fermion states must be either massless or
 degenerate in parity. It is evident that baryons  (particularly, the nucleon) do not
 possess  such properties. This means that the chiral symmetry of the QCD Lagrangian is
 spontaneously broken. According to the Goldstone theorem the spontaneous breaking of
 symmetry leads to appearance of massless particles in the spectrum of physical states -
 Goldstone bosons. (The proof of the Goldstone theorem for the case of QCD is given in
 [5].) In QCD Goldstone bosons  can be identified with the triplet of $\pi$-mesons
 in the limit $m_u,m_d \to 0,~m_s \not= 0$ (SU(2) symmetry) or with the octet of
 pseudoscalar mesons ($\pi, K, \eta)$ in the limit  $m_u, m_d ,m_s \to 0$ (SU(3)
 symmetry).

\vspace{7mm}

{\bf \large 2. The origin of baryon masses}

\bigskip

As was already said, there are two facts clearly  indicating on the spontaneous violation
of chiral  symmetry in QCD:\\
1) the existence of quark condensate and its typical hadronic scale;\\
2) large baryon (particularely, proton) masses.

The question arises: are these phenomena directly connected ? At first sight it seems
that it is  not. For dimensional grounds we can write
\be m^3 = - c\langle 0 \mid \bar{q} q \mid 0 \rangle, \label{3}\ee where $m$ the proton
mass and $c$ some numerical constant. The substitution of the value of proton mass and
of the numerical value (\ref{2}) in (\ref{3}) gives: $c\approx 50$ -- an unreasonable large
number. In fact the relation of the form (\ref{3}) takes place indeed and the approximate
formula reads [8]
\be m^3 = -2(2\pi)^2 \langle 0 \mid \bar{q}q \mid 0 \rangle \label{4}\ee The formula
(\ref{4}) has no adjustable parameters and its accuracy is about 15\%.

Let us now dwell on the derivation of a more precise expression for proton mass in QCD in
terms of the v.e.v.'s of various operators. It is used the Operator Product Expansion
(OPE) method [9], intensely exploited by Shifman, Vainshtein and Zakharov (SVZ) [10] in
investigation of the properties of QCD and determination of the meson masses. Define the
polarization operator:
\be \Pi(p) =i \int d^4 x e^{ipx} \langle 0 \mid T\{\eta (x),~ \bar{\eta}(0)\} \mid 0
\rangle,\label{5}\ee where $\eta(x)$ is three-quark current with proton quantum numbers
built from $u$ and $d$ quark fields. The explicit form  of the current $\eta(x)$ is:
\be \eta(x) =(u^a(x) C\gamma_{\mu} u^b(x))\gamma_5 \gamma_{\mu} d^c(x)
\varepsilon^{abc},\label{6}\ee where $a,b,c$ -- are the  colour indeces,
$\varepsilon^{abc}$ is the unit antisymmetric tensor, $C$ -- is charge conjugation
matrix. (The argumentation, why this form of the current is the most suitable for
determination of proton mass and other proton properties, was given in [8],[11].) The
general form of $\Pi(p)$ is
\be \Pi(p) =\not \!{p} \Pi_1(p^2) +\Pi_2(p^2),\label{7}\ee where $\not\!{p}=
p_{\mu}\gamma_{\mu}$. The first term in (\ref{7}) conserves chirality, while the second
violates chirality. The OPE is written separately for chirality conserving and chirality
violating structures:
\be \Pi(p^2) =\sum_n C^{(i)}_n (p^2) O^{(i)}_n (0),~~~i =1,2. \label{8}\ee In (\ref{8})
$O^{(i)}_n(0)$ are v.e.v.'s of  various operators, $C^{(i)}_n(p^2)$ -- are the
coefficient functions. OPE (\ref{8}) is valid at large negative $p^2$, more precisely at
$p^2 < 0, \mid p^2 \mid R^2_c \gg 1$, where $R_c$ is the confinement radius.

Consider first the chirality conserving structure $\Pi_1(p^2)$, perform the OPE and
classify the operators according to their dimensions $d_n$. The operator of dimension 0
is the unit operator, $O_1 =1$. Its contribution is proportional to $p^4\ln p^2$. (The
polynomical terms are disregarded, they are not contributing to the sum rules below.) The
next term of OPE has the dimension $d=4$ and is proportional to gluon condensate, $O_4
=(\alpha_s/\pi) \langle 0\mid G^n_{\mu\nu} G^n_{\mu\nu} \mid 0 \rangle$. (The quark
masses are neglected.) The next in dimension is the $d=6$ operator, corresponding to the
v.e.v. of the product of 4 quark fields, $O_6 = \langle 0 \mid \bar{q} \Gamma q \cdot
\bar{q} \Gamma q \mid 0\rangle,~q=u,d$. The contribution of this operator has no loop
integration (Fig.1c), $C_6 \sim 1/p^2$. The contribution of the other  $d=6$ operator,
proportional to the product of three  gluonic fields has two loop integration and is
strongly suppressed numerically. The corresponding diagrams (up to $d=8$) are shown in
Fig.1.
\begin{figure}[h]
\hspace{16mm} \epsfxsize=11.0cm \epsfbox{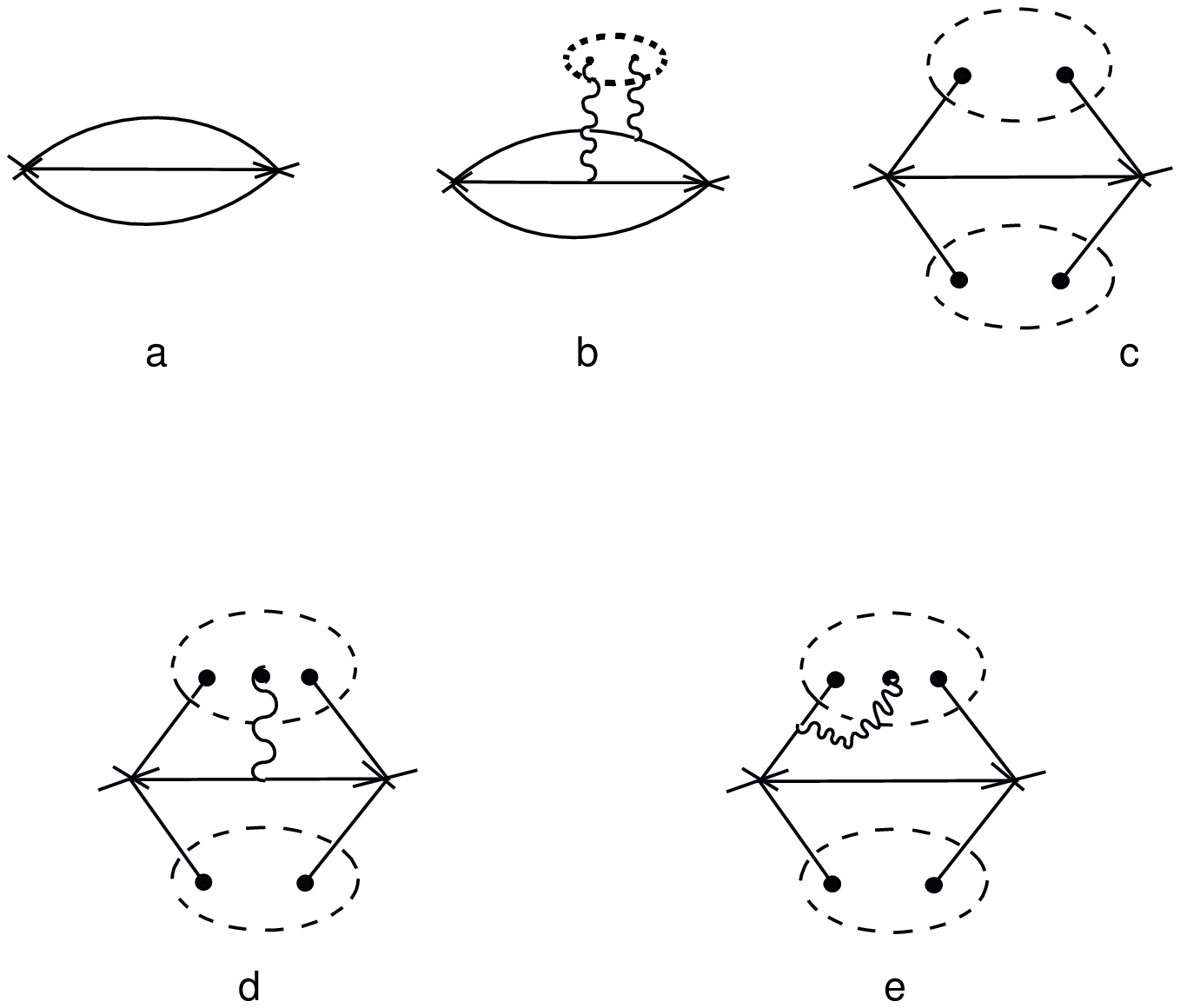}

{\bf Fig. 1.} Feynman diagrams for $\Pi_1(p^2)$ -- the chirality conserving part of the
polarization operator. The solid lines correspond to quarks,  wavy lines to gluons, dots
outlined by dashed lines stand for the mean vacuum values of the field operators, crosses
for interactions with the external current.
\end{figure}
The construction of OPE  for chirality violating structure $\Pi_2(p^2)$ is similar. The
lowest dimension operator is the quark condensate, $O_3 = \langle 0 \mid \bar{q} q\mid 0
\rangle$, $C_3^{(2)} \sim p^2 \ln p^2$. The next in dimension operator has $d=5$ and its
v.e.v. has to the form:
\be O_5 =-g \langle 0 \mid \bar{q} \sigma_{\mu\nu} \frac{\lambda^n}{2} G^n_{\mu\nu} q
\mid 0 \rangle \equiv m^2_0 \langle 0 \mid \bar{q}q \mid 0 \rangle, \label{9}\ee $C_5
\sim \ln p^2$. The corresponding diagrams are presented in Fig.2 (up to dimension $d=9$).
\begin{figure}[h]
\hspace{28mm} \epsfxsize=10.0cm \epsfbox{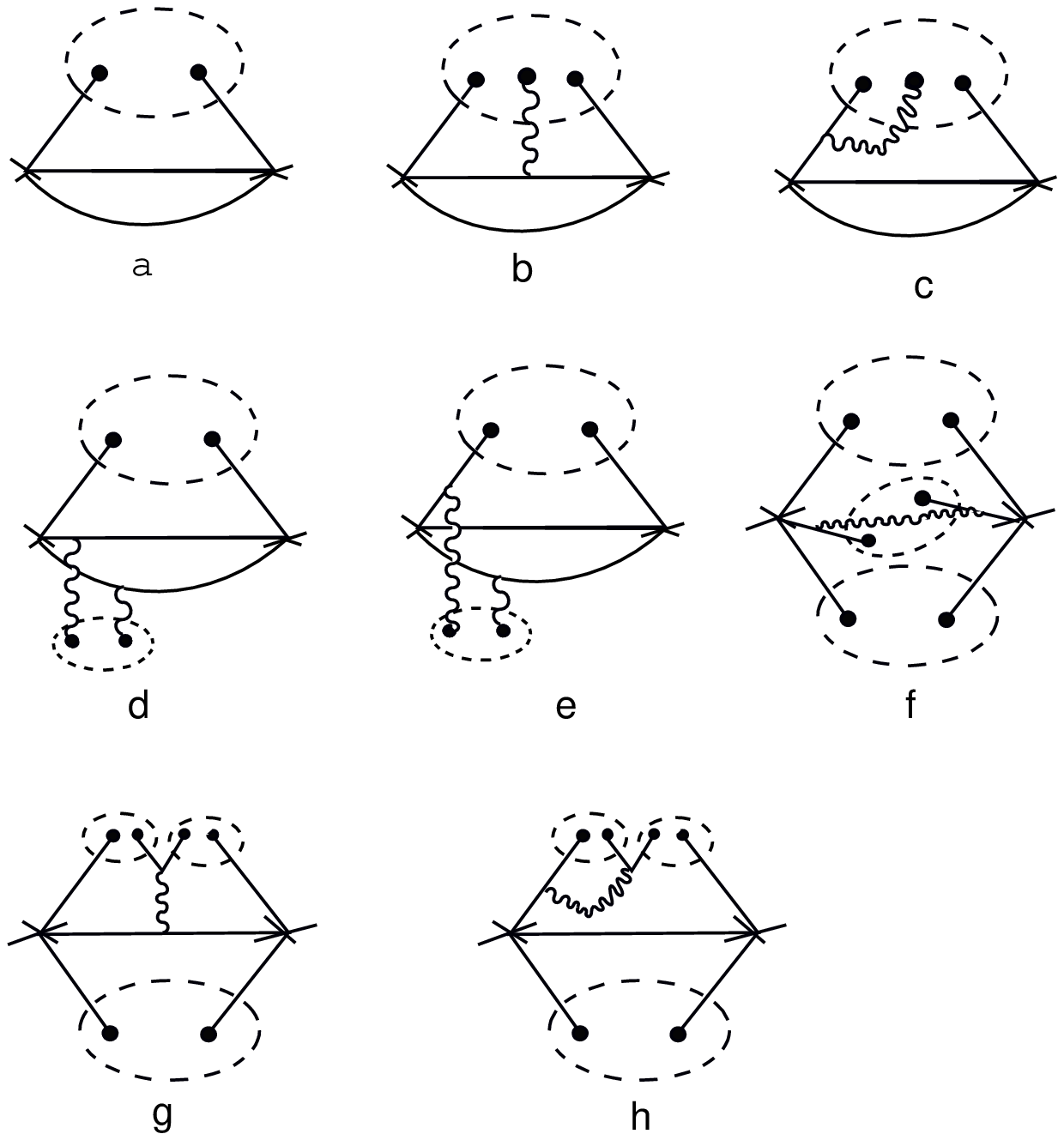}

 {\bf Fig. 2.} Feynman diagrams for $\Pi_2(p^2)$.
  The notation is the same as in Fig.1.
\end{figure}
In order to connect the polarization operator, calculated in terms of quark and gluon
condensates via OPE with the same polarization operator expressed in terms of hadronic
variables $\Pi^{phys}(p^2)$ use the dispersion relation  representation:
\be \Pi^{phys}_i (p^2) =\frac{1}{\pi} \int\limits^{\infty}_0 \frac{Im \Pi^{phys}_i
(s)}{s-p^2} ds + \mbox{subtractions}\label{10}\ee In the right hand side of (\ref{10}) the
model of hadronic spectrum is used:
\be Im \Pi^{phys}_i (s) = \mbox{resonance} + \mbox{continuum} \label{11}\ee It is
assumed, that the resonance is separated from continuum by the gap, the continuum is
equal to the  asymptotics of $\Pi_i(p^2)$ and starts at continuum threshold  $s_0$. The
equality
\be \Pi_i(p^2) = \Pi_i^{phys} (p^2) \label{12}\ee is the desired QCD sum rule [10].

In case of proton the resonance contribution is equal:
\be Im \Pi(p) = (\not \!{p} + m)\delta(p^2 -m^2)\lambda^2, \label{13}\ee  where $\lambda$
is given by:
\be \langle 0 \mid \eta \mid p \rangle =\lambda v_p \label{14}\ee and $v_p$ is the proton
spinor. Eq.(\ref{12}) is not well defined because of divergences at the left hand side
and subtraction terms at the right hand side. In order to kill both and improve the
convergence of OPE and dispersion relation representation SVZ suggested  to apply the
Borel transformation  to Eq.(\ref{12}). Put $Q^2 =-p^2$. Borel transformation of the
function $f(Q^2)$ is defined by:
\be ~~~{\cal{B}} f(Q^2)~ =~\lim_{n\to \infty} \frac{(Q^2)^{n+1}}{n!} \biggl (
-\frac{d}{dQ^2}\biggr )^n f(Q^2). \label{15}\ee {\scriptsize
$$\begin{array}{l}
Q^2 \to \infty\\
Q^2/n = M^2 = \mbox{Const.}
\end{array}
$$}
If $f(Q^2)$ can be represent by dispersion relation --
\be f(Q^2) =\frac{1}{\pi} \int \frac{Im f(s)}{s+Q^2} ds, \label{16}\ee then
\be {\cal{B}}_{M^2} f(Q^2) =\frac{1}{\pi} \int e^{-s/M^2} Im f(s) ds.\ee Borel
transformation improves the convergence of OPE, since
\be {\cal{B}}_{M^2} \frac{1}{(Q^2)^n} = \frac{1}{(M^2)^{n-1}(n-1)!}\label{18}\ee
Collecting all terms of OPE, represented by diagrams of Fig.'s 1 and 2 and applying the
Borel transformation  to (\ref{12}) we get the sum rules for photon polarization operator
[8],[12],[13]
$$ M^6E_2(M)L^{-4/9}c_0(M) + \frac{1}{4} b M^2E_0(M) L^{-4/9}+
\frac{4}{3} a^2_{\bar{q}q}c_1(M)  -\frac{1}{3} a^2_{\bar{q}q} \frac{m^2_0}{M^2}$$ \be =\
\tilde{\lambda}^2_N \exp \biggl (-\frac{m^2}{M^2}\biggr ), \label{19} \ee
\be 2a_{\bar{q}q} M^4 E_1 (M)  c_2(M) + \frac{272}{81} \frac{\alpha_s(M)}{\pi}
\frac{a^3_{\bar{q}q}}{M^2} -\frac{1}{12} a_{\bar{q}q}b = m \tilde{\lambda}^2_N \exp
\biggl (-\frac{m^2}{M^2}\biggr ).\label{20} \ee Here
\be a_{\bar{q}q}\ =\ -(2\pi)^2 \langle 0\mid \bar{q}q\mid 0 \rangle = 0.65~\mbox{GeV}^3,
 \label{21}\ee
\be
 b\ =\
(2\pi)^2 \langle0\mid \frac{\alpha_s}\pi G^2_{\mu\nu}\mid 0 \rangle, \label{22}\ee
\be
 L\ =\
\frac{\alpha_s(\mu^2)}{\alpha_s(M^2)}\,, \label{23}\ee
\be
 E_n(M)\ =\ \frac1{n!}
\int\limits_0^{s_0/M^2} z^ne^{-z} dz\,. \label{24} \ee
\be \tilde{\lambda}^2 = 2(2\pi)^2 \lambda^2,\label{25}\ee $\mu$ is the normalization
point and $c_0,c_1,c_2$ are $\alpha_s$ corrections [14],[15],[16],[17]. (In the
derivation of Eq.'(\ref{19}),(\ref{20}) the factorization hypothesis  [10] for
contributions of operators of higher dimensions $(d \geq 6)$ was assumed: the saturation
of such contribution by vacuum intermediate state. This assumption is legitimate at large
number of colours $N_c$ and the corrections to factorized formulae are of order $1/N^2_c \sim 10\%$.) As will be
shown below, Eq.'s (\ref{19}),(\ref{20}) are valid at $M^2=0.9-1.5$ GeV$^2$. Let us
perform the rough approximation: neglect all higher order terms of OPE and continuum
contribution ($s_0 \to \infty, ~E_i=1$), as well as $\alpha_s$-corrections ($L=1,
c_k=1$). Put $M^2=m^2$ and divide (\ref{20}) by (\ref{19}). We get Eq.(\ref{4}) presented
above. The substitution of the numerical value of quark condensate (\ref{2}) gives:
$m=1.09$ GeV in comparison with the experimental value $m_{\exp} = 0.94$ GeV. Go now to a
more exact treatment of Eq.'s (\ref{19}),(\ref{20}). The values of $m$, found as the
ratio of (\ref{20}) to (\ref{19}) and $\tilde{\lambda}^2$ from (\ref{19}) and (\ref{20})
at $m=m_{\exp} =0.94$ GeV are plotted in Fig.3 as functions of $M^2$. (For the values of
parameters -- see [7].)
\begin{figure}[h]
\hspace{15mm} \epsfxsize=13.0cm \epsfbox{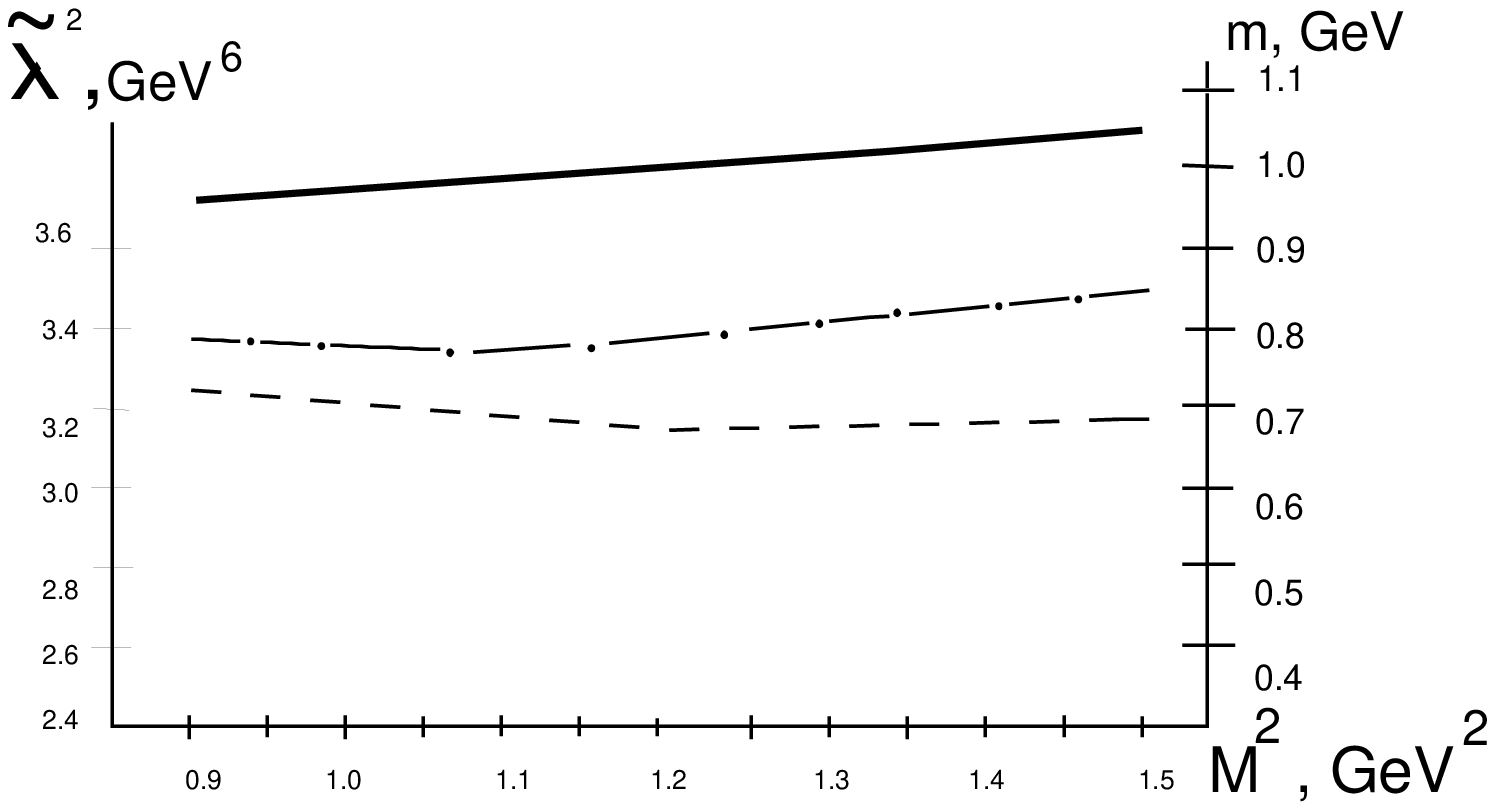}

{\bf Fig. 3.} The proton mass sum rules, Eqs. (\ref{19}) and (\ref{20}). The dashed and
dash--dotted curves give $\tilde{\lambda}^2$, determined from (\ref{19}) and (\ref{20})
respectively, using the experimental value of $m$ (left scale).  The solid line gives $m$
as the ratio of (\ref{20}) to (\ref{19}).
\end{figure}

As is seen from Fig.3 the $M^2$-dependence of $m$ and $\tilde{\lambda}^2$ is  very weak
at $0.9 < M^2 < 1.5$ GeV and the values of $\tilde{\lambda}^2$ found from (\ref{19}) and
(\ref{20}) differ less than 5\% at $M^2\approx 1$ GeV$^2$. The final result of the proton
mass calculation is [8],[12],[13] (see also [7],[18]):
\be m=0.98 \pm 0.10~ \mbox{GeV}\label{26}\ee
\be \tilde{\lambda}^2 = 3.2 \pm 0.6~ \mbox{GeV}^6 \label{27}\ee The masses of hyperons
and various baryon resonances were calculated in a similar way [8],[19] with a good
coincidence  with experiment. Particularly, the mass of $\Delta$-isobar was found to be
equal to
\be m_{\Delta} =1.30 \pm 0.18~\mbox{GeV}~~(m_{\Delta~\exp}=1.23~\mbox{GeV})\label{28}\ee
The conclusion from all the said above is: the appearance of baryon masses and their
numerical values is caused by chiral symmetry violation in QCD vacuum.

\vspace{7mm}

{\bf \large 3. Baryon magnetic moments}

\bigskip

The vacuum in QCD can be considered as continous medium. Under the influence of external
electromagnetic field $F_{\mu\nu}$ quark pairs in the vacuum are polarized: it appears to
be induced by the field vacuum expectation value [20],[13]
\be \langle 0 \mid \bar{q} \sigma_{\mu\nu} q \mid 0 \rangle_F = \chi_q \langle 0\mid
\bar{q} q \mid 0 \rangle F_{\mu\nu},~~~q=u,d,s.\label{29}\ee We restrict ourselves by
consideration of the constant electromagnetic field, the electric charge $e=\sqrt{4\pi
\alpha_{em}}$ is included in $F_{\mu\nu}$. The factor $\langle 0 \mid \bar{q} q \mid
0\rangle$ is separated in (\ref{29}),  since $\langle 0 \bar{q}\sigma_{\mu\nu} q \mid 0
\rangle$  is violating chirality. This phenomenon  is similar  to paramagnetism or
diamagnetism in matter. It can be shown [20],[13] that in a good approximation
\be \chi_q = e_q\chi\label{30}\ee where $e_q$ is the $q$-quark charge. The arguments are
the following. The appearance of the charge of some other quark $q'$,  not coinciding  with
$q,~q'\not=q$ is caused by the diagram of Fig.4
\begin{figure}[tb]
\hspace{45mm} \epsfig{file=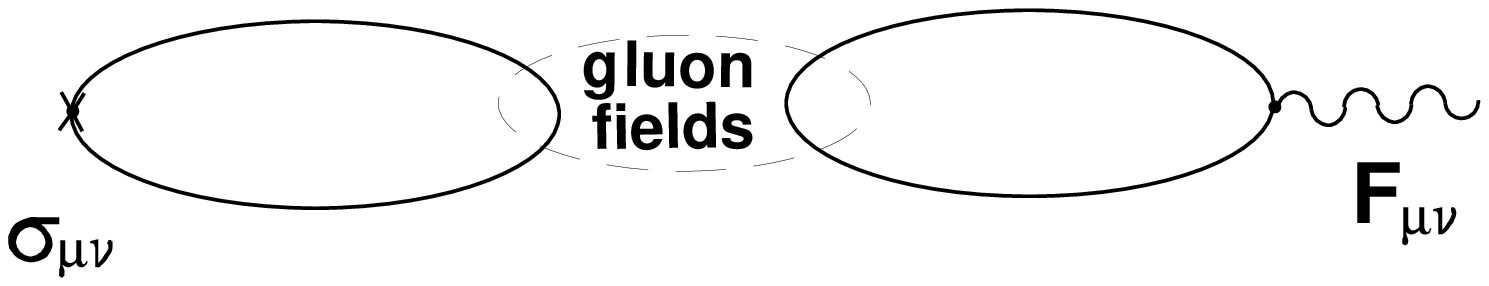, width=73mm}

\vspace{3mm} {\bf Fig. 4.} The diagrams corresponding to quark pair mixing, $\bar{q}q\to
\bar{q}^{\prime}q^{\prime},$ and resulting in deviations from Eq.(\ref{30}).
\end{figure}
However, the diagram of Fig.4 is zero in any order of perturbation theory because of
chirality conservation. Chirality violation might appear due to instantons, but for
massless quarks the loop in Fig.4 vanishes -- its $G$-parity in colour space is negative,
$G_{colour} = -1$  [21]. The amplitude of Fig.4 has some resemblance to $\varphi-\omega$
mixing. Therefore the experimental smallness of $\varphi-\omega$ mixing is also an
argument in favour of (\ref{20}). The universal constant $\chi$ is called the quark
condensate magnetic susceptibility. For consideration of the problem add to QCD
Lagrangian the term, corresponding to interaction with electromagnetic field
\be L' =\int d^4 x j^{el}_{\mu} A^{el}_{\mu} = \int d^4 x j^{el}_{\mu} \frac{1}{2}
x_{\nu} F_{\nu \mu}.\label{31}\ee The  Fock-Schwinger gauge for e.m. field: $x_{\mu}
A_{\mu}^{el}(x) =0$ is used. The polarization  operator in the linear approximation in
$F_{\mu\nu}$ has the form:
\be \Pi(p)_F = i\int d^4 x e^{ipx} \langle 0 \mid T\{ \eta(x), ~\bar{\eta}(x)\} \mid
0\rangle_F = \Pi^{(0)} (p) +\Pi^{(1)}_{\mu\nu} (p) F_{\mu\nu}.\label{32}\ee We are
interested in $\Pi^{(1)}_{\mu\nu}(p)$. Perform OPE and classify the v.e.v. of operators
according to their dimensions. The operator of the lowest dimension with $d=2$ is
$F_{\mu\nu}$ itself. The  next, with $d=3$, and for this reason the most important, is
$\langle 0 \mid \bar{q} \sigma_{\mu\nu} q \mid 0 \rangle_F$. There are two v.e.v.'s
operators of dimension 5
\be g \langle 0\mid \bar{q} \frac{1}{2} \lambda^n G^n_{\mu\nu} q \mid 0\rangle_F
=\kappa_q F_{\mu\nu} \langle 0 \mid \bar{q}q\mid 0 \rangle \label{33}\ee
\be -ig\varepsilon_{\mu\nu\lambda\sigma} \langle 0 \mid \bar{q} \gamma_5 \frac{1}{2}
\lambda^n G^n_{\lambda\sigma}q \mid 0\rangle_F =\xi_q F_{\mu\nu} \langle 0 \mid \bar{q} q
\mid 0 \rangle.\label{34}\ee By analogy with (\ref{20}) $\kappa_q$ and $\xi_q$ are
proportional to quark charge:
\be \kappa_q =e_q\kappa_1,~~~\xi_q =e_q \xi\label{35}\ee Among 6-dimensional operator the
vacuum everage
\be \langle 0 \mid \bar{q}q \mid 0 \rangle \langle 0\mid \bar{q} \sigma_{\mu\nu} q\mid 0
\rangle_F\label{36}\ee is accounted  for (within the framework  of the factorization
hypothesis). In order to obtain the sum rules for polarization operator in the external
electromagnetic field we need its  dispersion relation represention. $\Pi_{\mu\nu}(p)_F$
corresponds  to three point function. So, generally, we shall start from consideration of
$\Pi_{\mu\nu} (q;p_1,p_2)$, where $q$ --is the momentum carried by electromagnetic field.
The spectral representation of any structurte function $\Gamma(p^2_1,p^2_2,q^2)$ of
$\Pi_{\mu\nu} (q;p_1,p_2)$ is given by the double dispersion relation in variables
$p^2_1, p^2_2$ at fixed $q^2 \leq 0$:
\be \Gamma(p^2_1,p^2_2; q^2) = \int\limits^{\infty}_0\int\limits^{\infty}_0
\frac{\rho(s_1,s_2,q^2)}{(s_1-p^2_1)(s_2 -p^2_2)} ds_1 ds_2 + P(p^2_1)f(p^2_2,
q^2)+P(p^2_2)f(p^2_1,q^2).\label{37}\ee In the final result the limit $q^2\to 0, p^2_1
\to p^2_2=p^2$ will be performed. The proton contribution corresponds to the term in
$\rho(s_1,s_2,q^2)$, proportional to $\delta(s_1-m^2)\delta(s_2-m^2)$ and the double pole
in $\Gamma$:
\be \Gamma(p^2) \sim \frac{1}{(p^2 -m^2)^2}.\label{38}\ee In Eq. (\ref{37}) the inelastic
transitions, represented by Fig.5, contribute to subtraction terms.
\begin{figure}[tb] \hspace{55mm}
\epsfig{file=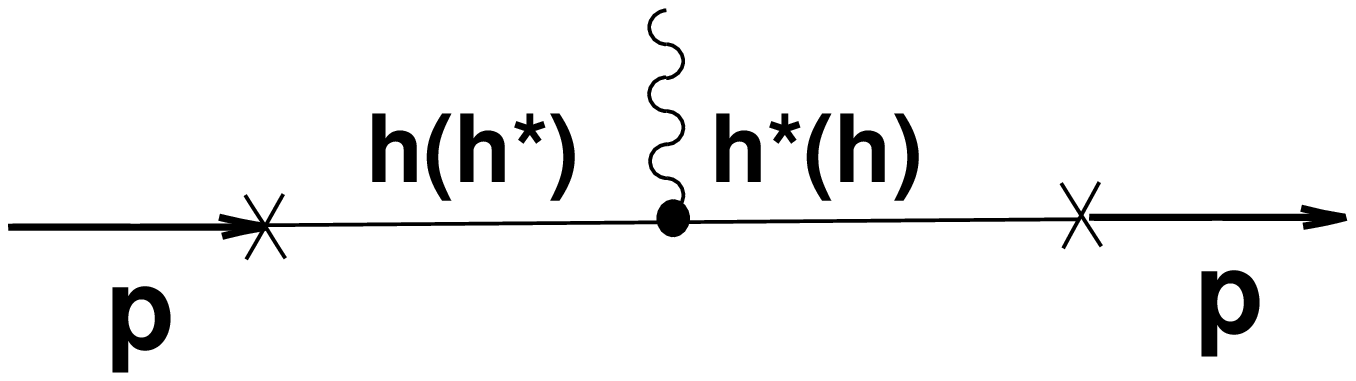, width=55mm}

\vspace{3mm} {\bf Fig. 5.} The schematical  representation of $h \longrightarrow h^*$
($h^*\longrightarrow h$) transitions in the external field.
\end{figure}
Their contribution is proportional to:
\be \frac{A}{(p^2-m^2)(p^2-m^{*2})},\label{39}\ee i.e. has a single pole at $p^2=m^2$.
There are three tensor structures of $\Pi_{\mu\nu}(p)$:
\be \not\!{p} \sigma_{\mu\nu} +\sigma_{\mu\nu}\hat{p},~~~i(p_{\mu}\gamma_{\nu}
-p_{\nu}\gamma_{\mu})\not\!{p},~~~\sigma_{\mu\nu}\label{40}\ee The first structure
conserves chirality, two last are violating chirality. The structure function at the last
structure is not suitable for obtaining the sum rules, because of a bad convergence of
OPE and large contribution of instantons [22]. For the first two structures Eq.(\ref{37})
in the limit $q^2\to 0, p^2_1\to p^2_2 \to p^2$ reduces to
\be \Gamma(p^2) =\int \frac{\rho(s)}{(s-p^2)} + \frac{A}{s-p^2}.\label{41}\ee (The
treatment of a more general case, which arises, e.g. if $\alpha_s$ -- corrections are
taken into account, can be performed using the method of [23].) I will not present here
the sum rules for the proton and neutron separately. (They are presented in [13]). It is
possible to obtain the combination of the  sum rules for proton and neutron, where all
unknown susceptibilities -- $\chi,\kappa$ and $\xi$ are excluded. These sum rules can be
represented in the form:
\be \mu_pe_d -\mu_ne_u +M^2(A_pe_d - A_n e_u) = \frac{4
a^2_{\bar{q}q}}{3\tilde{\lambda}^2} e^{m^2/M^2} (e^2_u - e^2_d) L^{4/9}, \label{42} \ee
\be \mu^a_p e_u - \mu_n e_d + M^2 (B_p e_u - B_n e_d) =\frac{4a_{\bar{q}q} m
M^2}{\tilde{\lambda}^2} e^{m^2/M^2} (e^2_u - e^2_d). \label{43} \ee Here the constants
$A$ and $B$ represent the contributions of inelastic transitions for the first and second
structures correspondingly, the indeces $p$ and $n$ mean proton and neutron, $\mu$ are
the magnetic moments (in nuclear magnetons) $\mu^a_p$ is the proton anomalous magnetic
moment.  In order to eliminate the unknown single-pole contributions still remaining on
the l.h.s. of Eqs. (\ref{42}),\,(\ref{43}) we apply the differential operator $1-M^2
\partial/\partial M^2$ to these equations and obtain
\be \mu_p e_d-\mu_n e_u = \frac{4a^2_{\bar{q}q}}{3\tilde{\lambda}^2_N} (e^2_u - e^2_d)
\biggl (1 - M^2 \frac{\partial}{\partial M^2} \biggr ) e^{m^2/M^2} L^{4/9}, \label{44}\ee
\be
 \mu_p e_u -\mu_ne_d = e_u +\frac{4a_{\bar{q}q}m}{\tilde{\lambda}^2_N} (e^2_u - e^2_d)
\biggl ( 1- M^2\frac{\partial}{\partial M^2}\biggr ) M^2 e^{m^2/M^2}. \label{45}\ee  The
magnetic moments $\mu_p$ and $\mu_n$ can be approximately determined by setting $M=m$,
disregarding anomalous dimensions and substituting for the residue $\tilde{\lambda}^2_N$
the value
\be \tilde\lambda^2_N\ =\ \frac{2a_{\bar{q}q}M^4}m e^{m^2/M^2}\biggl\vert _{M^2=m^2}\,,
\label{46} \ee
 which follows from the mass sum rules (\ref{20}) neglecting
 both anomalous dimensions and continuum contributions. Solving in
 this approximation Eqs. (\ref{44}), (\ref{45}) we arrive at
 the elegant results:
\be \mu_p\ =\ \frac{8}{3}\biggl (1+\frac16\frac{a_{\bar{q}q}}{m^3}\biggr), \label{47} \ee
\be \mu_n\ =\ -\frac43 \biggl(1+\frac23\frac{a_{\bar{q}q}}{m^3}\biggr). \label{48} \ee
Numerically, at $a_{\bar{q}q}=0.65\rm\,GeV^3$  we get from
(\ref{47}),(\ref{48}) $\mu_p=3.01, \mu_n =-2.03$ in comparison with the experimental
values $\mu_p=2.79, \mu_n=-1.91$. In a more rigorous treatment the study of the
$M^2$-dependence of Eqs. (\ref{44}),(\ref{45}) in the confidence interval
$0.9<M^2<1.3\rm\,GeV^2$ gives as the best fit the values
\be \mu_p=2.7, \qquad \mu_n=-1.7 \label{49} \ee with an estimated error of about 10\%.
The proton and neutron  magnetic moment were also calculated in [24],[25]. In [24] the
sum rule for the chirality conserving structure was only used and the susceptibility
$\chi$ was estimated basing on the vector dominance model. In [25] it was shown, that
gluon condensate contribution, neglected in the above consideration is very small indeed
(less than 1\%). The hyperon magnetic moments were calculated in a similar way [26]. The
results are presented in the Table.

\begin{center}
{\bf Table } \vspace{3mm}

{\bf Magnetic moments of the baryon octet.}

\vspace{3mm}

\begin{tabular}{|l|l|l|l|l|l|l|l|l|l|} \hline
 & ~~$p$ &~~$n$ & ~$\Sigma^+$ & ~$\Sigma^0$ & ~~$\Sigma^-$ &
 ~~$\Xi^0$& ~~$\Xi^-$ & ~~$\Lambda$ & ~$\Sigma\Lambda$ \\ \hline
 sum  rules & 2.70 & -1.70 & 2.70 & 0.79 & -1.12 & -1.65 &
 -1.05 & -0.72$^{b)}$ & 1.54$^{b)}$ \\ \hline quark
 model & 2.79$^{a)}$ & -1.91$^{a)}$ & 2.67 & 0.78 & -1.09 & -1.44 &
 -0.49 & -0.61$^{a)}$ & 1.63 \\ \hline experiment &  2.79 & -1.91
 & 2.46 &  ~~-- & -1.16 & -1.25 & -0.65 & -0.61 & 1.61 \\ \hline
 \end{tabular}
\end{center}
 a) Input data;\\
b) Approximate value, calculated on the basis of $SU(3)$ relations.

\vspace{3mm}

\noindent Within the limits of the expected theoretical errors (10--15\%) the results of
the sum rule calculations are in agreement with the data. The exceptional case is the
$\Xi$ hyperon, where the difference between theory and experiment is larger. The latter
can be addressed to a significant $M^2$ dependence of the sum rules for the $\Xi$-mass
and magnetic moments, which in turn may be related to a larger role of $m_s$-corrections.
In conclusion, it must  be emphasized, that no new parameters, besides those found in the
calculations of baryon masses, enter the above used sum rules  for baryon magnetic
moments. The quark condensate magnetic susceptibility  was found with the help of the
special sum rule [27],[28],[29]. The most precise result is [29]:
\be \chi_{1~GeV}= -(3.15 \pm 0.3)~\mbox{GeV}^{-2}.\label{50}\ee

\vspace{7mm}

{\bf \large 4. The nucleon coupling constants with axial current}

\bigskip

The nucleon coupling constant with the isovector axial current $g_A$ determines the
$\beta$-decay rate of neutron, the nucleon coupling constant $g^{(8)}_A$ with the
8-component of octet axial current determines (together with $g_A$) the $\beta$-decays of
hyperon. Therefore their knowledge is very important. I will not dwell on the theoretical
formulae for these quantities -- their derivation is based on the method, similar to that
used in the calculation of baryon magnetic moments -- and presents here only the final
results:
\be g_A^{theor} = 1.24 \pm 0.05~[30],[31],[32]~~(\exp.1.269 \pm 0.003~ [6])\label{51}\ee
\be g_A^{(8)theor} = 0.45 \pm 0.15~[33],[32]~~(\exp.0.59 \pm 0.02~ [33])\label{52}\ee When
calculating  $g_A^{theor}$ and $g_A^{(8)theor}$ the sum rule for nucleon mass, as well
as the numerical value of nucleon coupling constant $\tilde{\lambda}^2$ with quark
current $\eta$ was exploited. The agreement with experiment is good, especially in the
case of $g_A$. (The large error in $g^{(8)~theor}_A$ arises from strong compensation of
the main term in OPE -- see [32].) In the limit of exact SU(3)-symmetry $g_A$ and
$g_A^{(8)}$ are related with $\beta$-decay axial coupling constants in the baryon octet
$F$ and $D$:
$$ g_A = F+ D$$
\be g^{(8)}_A =3F - D\label{53}\ee From (\ref{52}),(\ref{53}) we have
\be F^{theor} =0.42 \pm 0.04; ~~D^{theor} = 0.82 \pm 0.08\label{54}\ee

Of a special interest is the nucleon coupling constant with singlet axial current
\be j^{(0)}_{\mu 5} =\bar{u} \gamma_{\mu} \gamma_5 u + \bar{d}\gamma_{\mu}\gamma_5 d
+\bar{s} \gamma_{\mu} \gamma_5 s\label{55}\ee Due to the Bjorken sum rule this coupling
constant is connected with the part of proton spin $\Sigma$, carried by quarks. (See [34]
for review and references). The singlet axial current is not conserved because of
anomaly. In order to find $\Sigma$ it was considered the sum rule for the topological
current
\be Q_5 =\frac{\alpha_s}{8\pi} G_{\mu\nu}\tilde{G}_{\mu\nu},~~~\tilde{G}_{\mu\nu} =
\frac{1}{2} \varepsilon_{\mu\nu\lambda\sigma} G_{\lambda\sigma}\label{56}\ee The result
is [32]
\be\Sigma =0.30 \pm 0.18 \label{57}\ee The value of $\Sigma$ (\ref{57}) agrees with the
data (although with large errors, the review on proton spin structure is given in the
book [35], see also [36]). This value agrees well with the vector based dominance model
[37],[38],[39] establishing the smooth connection of the integrals of polarized structure
function
\be \int g_1(x,Q^2) dx \label{58}\ee at $Q^2=0$ and high $Q^2$. (The former is given by
Gerasimov-Drell-Hearn sum rule [40],[41] and expressed through nucleon magnetic moments.)

\vspace{7mm}

{\bf \large 5. Conclusion}

\bigskip

1. It was demonstrated  that the origin of baryon masses is the spontaneous violation of
chiral symmetry in QCD vacuum -- the existance of quark condensate.

2. Basing on this statement the proton mass was calculated with accuracy $\sim 10\%$, as
well as hyperon masses and masses of various baryon resonances.

3. It was shown, that the same phenomenon -- the existance of quark condensate and its
polarizability  in external electromagnetic field is responsible  for the magnitudes of
the proton, neutron and hyperons magnetic moments.

4.  The consequence of this is, that the properties  of baryons are determined by the
properties of QCD vacuum and are weakly related to the  structure of theory  at small
distances (existance of Higgs boson, its interaction with quark etc.)

This work was supported in part by  RFBR grant 06-02-16905a and
the funds from EC to the project ``Study of Strongly Interacting
Matter'' under contract No. R113-CT-2004-506078.

\vspace{1cm}

\centerline{\bf  REFERENCES}

\bigskip

\noindent ~1. J. Gasser, H. Leutwyler, Nucl. Phys.  {\bf B94}, 269 (1975).

\vspace{2mm} \noindent
 ~2. S. Weinberg, in A.Festschrift for I.I.Rabi, ed. L.Motz, Trans. New
York Acad. Sci.,

Ser.II, 1978,  v.38. p.185.

\vspace{2mm} \noindent ~3.  B. L. Ioffe, Int. J.  Mod. Phys. A {\bf 21}, 6249 (2006).

\vspace{2mm} \noindent ~4.  M. Gell-Mann, R. J. Oakes, B. Renner, Phys. Rev. {\bf 175},
2195 (1968).

\vspace{2mm} \noindent ~5. B. L. Ioffe, Phys. Usp. {\bf 44}, 1211 (2001); Usp. Fiz. Nauk.
{\bf 44}, 1273 (2001).

\vspace{2mm} \noindent ~6.  W.-M.  Yao {\it et al.}, Review of Particle Physics, J. Phys.
G. {\bf 33}, 1 (2006).

\vspace{2mm} \noindent ~7. B. L. Ioffe, Prog. Part. Nucl. Phys. {\bf 56}, 232 (2006).

\vspace{2mm} \noindent ~8.  B. L. Ioffe, Nucl. Phys. B {\bf 188}, 317 (1981); Errata {\bf
B191}, 191.

\vspace{2mm} \noindent ~9.  K. Wilson, Phys. Rev. {\bf 179}, 1499 (1969).

\vspace{2mm} \noindent 10. M. A. Shifman, A. I. Vainshtein, V. I. Zakharov, Nucl. Phys.
{\bf B147}, 385,448 (1979).

\vspace{2mm} \noindent 11. B. L. Ioffe, Z. Phys. C. {\bf 18}, 67 (1983).

\vspace{2mm} \noindent 12. V. M.  Belyaev, B. L. Ioffe, Sov. Phys. JETP {\bf 56}, 493
(1982).

\vspace{2mm} \noindent 13.  B. L. Ioffe, A. V. Smilga, Nucl. Phys. B. {\bf 232},
 109 (1984).

\vspace{2mm} \noindent 14. M.  Jamin, Z. Phys. C. {\bf 37}, 635 (1988).

\vspace{2mm} \noindent 15. M. Jamin, Dissertation thesis, Heidelberg preprint
HD-THEP-88-19 (1988).

\vspace{2mm} \noindent 16.  A. A. Ovchinnikov, A. A. Pivovarov, L. R. Surguladze, Sov. J.
Nucl. Phys. {\bf 48}, 358

(1988).

\vspace{2mm} \noindent 17.  A. G. Oganesian,  hep-ph/0308289.

\vspace{2mm} \noindent  18. V. A. Sadovnikova, E. G. Drukarev, M. G. Ryskin, Phys. Rev.
{\bf D72}, 114015 (2005).

\vspace{2mm} \noindent 19.  V. M. Belyaev, B. L. Ioffe, Sov. Phys. JETP {\bf 57}, 716
(1983).

\vspace{2mm} \noindent 20. B. L. Ioffe, A. V. Smilga,  JETP Lett. {\bf 37},
 298 (1983).

\vspace{2mm} \noindent  21.  B. V. Geshkenbein, B. L. Ioffe, Nucl. Phys. {\bf B166}, 340
(1980).

\vspace{2mm} \noindent 22. M. Aw, M. K. Banerjee, H. Forkel, Phys. Lett.
{\bf 454B}, 147 (1999).

\vspace{2mm} \noindent 23.  B. L. Ioffe, Phys. At. Nucl. {\bf 58},  1408 (1995).

\vspace{2mm} \noindent  24.  I. I. Balitsky, A. V. Yung, Phys. Lett.
{\bf 129B}, 384 (1983).

\vspace{2mm} \noindent 25.  S. L. Wilson, J. Pasupathy, C. H. Chiu, Phys. Rev. {\bf D36},
1442 (1987).

\vspace{2mm} \noindent  26.  B. L. Ioffe, A. V. Smilga, Phys. Lett. {\bf 133B}, 436 (1983).

\vspace{2mm} \noindent 27. V. M. Belyaev, I. I. Kogan, Yad. Fiz. {\bf 40}, 1035 (1984).

\vspace{2mm} \noindent 28. I. I. Balitsky, A. V. Kolesnichenko, A. V. Yung, Sov. J. Nucl.
Phys. {\bf 41}, 178 (1985).

\vspace{2mm} \noindent 29. P. Ball, V. M. Braun, N. Kivel, Nucl. Phys.
{\bf  B649}, 263 (2003).

\vspace{2mm} \noindent 30.  V. M. Belyaev, Ya. I. Kogan,  JETP Lett. {\bf 37}, 730
(1983).

\vspace{2mm} \noindent  31. V. M. Belyaev, Ya. I. Kogan, Phys. Lett. {\bf 136B}, 273 (1984).

\vspace{2mm} \noindent  32. B. L. Ioffe, A. G. Oganesian, Phys. Rev. {\bf D57}, R6590 (1998).

\vspace{2mm} \noindent 33.  V. M. Belyaev, B. L. Ioffe, Ya. I. Kogan, Phys. Lett. {\bf 151B},
 290 (1985).

\vspace{2mm} \noindent  34. B. L. Ioffe, {\it Intern. School of Nucleon Structure,
1-st Course: The Spin Structure}

{\it of the Nucleon, Erice--Sicily},
eds. Frois B., Hughes V. (New York Plenum, 1997),

p. 215, hep-ph/9511401.

\vspace{2mm} \noindent  35.  S.D.  Bass,  {\it The Spin Structure of the Proton}
(World Sci., 2008).

\vspace{2mm} \noindent  36. S.D.Bass, B.L.Ioffe, N.N.Nikolaev, A.W.Thomas, J. Moscow.
Phys. Soc.

{\bf 1}, 317 (1991).

\vspace{2mm} \noindent  37.  M. Anselmino, B. L. Ioffe, E. Leader, Sov. J. Nucl. Phys.
{\bf 49}, 136 (1989);

Yad. Fiz. {\bf 49}, 214 (1989).

\vspace{2mm} \noindent  38.  V. D. Burkert, B. L. Ioffe, Phys. Lett.,  {\bf B296},
 223 (1992).

\vspace{2mm} \noindent  39.  B. L. Ioffe, Phys. At. Nucl. {\bf 60}, 1707 (1997).

\vspace{2mm} \noindent  40.  S. B. Gerasimov, Sov. J. Nucl. Phys. {\bf 2}, 930 (1966).

\vspace{2mm} \noindent 41.  S. D. Drell, A. C. Hearn, Phys. Rev. Lett. {\bf 16}, 908 (1966).

\end{document}